\documentclass[aps,pre,reprint,superscriptaddress,showpacs]{revtex4-1}
\usepackage{amsthm}
\usepackage{amssymb}
\usepackage[intlimits]{amsmath}
\usepackage{url}
\usepackage{hyperref}
\usepackage{graphicx}

\begin{document}



\title{ Universal asymptotic behavior in nonlinear systems driven by a two-frequency forcing}


\author{Jes\'us Casado-Pascual}
\email[]{jcasado@us.es}
\affiliation{F\'{\i}sica Te\'orica, Universidad de Sevilla, Apartado de Correos 1065, 41080 Sevilla, Spain}
\author{David Cubero}
\email[]{dcubero@us.es}
\affiliation{Departamento de F\'{\i}sica Aplicada I, EUP, Universidad de Sevilla, Calle Virgen de \'Africa 7, 41011 Sevilla, Spain}
\author{Ferruccio Renzoni}
\email[]{f.renzoni@ucl.ac.uk}
\affiliation{Department of Physics and Astronomy, University College London, Gower Street, London WC1E 6BT, United Kingdom}

\date{\today}

\begin{abstract}
We examine the time-dependent behavior of a nonlinear system driven by a two-frequency forcing. By using a
non-perturbative approach, we are able to derive an asymptotic expression,
valid in the long-time limit, for the time average of the output variable which describes the response of the system.  We identify several universal features of the asymptotic response of the system, which are independent of the details of the model. In particular, we determine an asymptotic expression for 
the width of the resonance observed by keeping one frequency fixed, and varying the other one. We show
that this width is smaller than the usually assumed Fourier width  by a 
factor determined by the two driving frequencies, and independent of the model system parameters.
Additional general features can also be identified depending on the specific symmetry properties of the system. Our results find
direct application in the study  of sub-Fourier signal processing with nonlinear systems.

\end{abstract}


\maketitle

\section{ Introduction}

Nonlinear systems driven by time-dependent forcings give rise to a wealth of intriguing phenomena, from chaos to the stochastic resonance phenomenon and the ratchet effect \cite{Reviews}. 
Given the complexity of the systems, a complete analysis of these phenomena is usually only possible via a numerical treatment, which 
is inherently model dependent.  A more general analysis, less model dependent, is usually possible only for the stationary state of the system, 
in the presence of system symmetries. These approaches thus do not easily allow the determination of the asymptotic scaling properties of 
the systems, whose identification is essential for the understanding of unique features of non-linear systems, as
the recently highlighted sub-Fourier dynamics which may open the possibility of sub-Fourier information processing \cite{lille, Optical_Trap3}. 

In this work, we study theoretically the time-dependent behavior of a generic nonlinear system driven by a two-frequency forcing. To specify the notation, we refer to a single-particle model, and assume that the average particle velocity is the quantity of interest for the study of the response of the  system. This is for example the case relevant to the study of directed
motion in driven systems. By using a nonperturbative approach, we are able to derive an asymptotic expression, valid in the long-time limit, for the average particle velocity. This expression allows us to identify several universal features of the asymptotic response
of the system, which are independent of the details of the model.
 Additional general features can also be identified depending on the 
specific symmetry properties of the system.

This paper is organized as follows. In Sec.~\ref{SOP}, we introduce the system of interest and state the problem under consideration. In Sec.~\ref{DOAP}, we
establish some general asymptotic properties of the nonlinear system, valid in the long-time limit. In Sec.~\ref{MSE}, we illustrate our theoretical results with the help of a one-dimensional model consisting of a Brownian particle, moving in a periodic potential, under the influence of a biharmonic force. We also investigate how the specific symmetries of the evolution equation affect our asymptotic results. Finally, in Sec.~\ref{Conclusions}, we present conclusions for the
main findings of our work.


\section{ Statement of the problem}
\label{SOP}

Let us consider a generic single-particle model in which the time evolution of the particle's position vector $\mathbf{r}(t)$  is described by a differential equation, 
which may be stochastic or deterministic depending on the presence or absence of noise sources.
From the solution of this differential equation with an appropriate set of initial conditions at an arbitrary time $t_0$, one 
can determine the particle's position vector at any later time $t_{\mathrm{f}}> t_0$. The average velocity between these two times is 
then defined as
\begin{equation}
\label{AVdefinition}
\mathbf{\bar{v}}^{t_0,t_{\mathrm{f}}}=\frac{1}{T_{\mathrm{e}}}\int_{t_0}^{t_{\mathrm{f}}}\mathrm{d}t\,\left\langle\dot{\mathbf{r}}(t)\right\rangle =\frac{\left\langle \mathbf{r}(t_\mathrm{f}) - \mathbf{r}(t_0)\right\rangle}{T_{\mathrm{e}}}\,,
\end{equation}
where $T_{\mathrm{e}}=t_{\mathrm{f}}-t_0$ is the elapsed time, the overdot represents time derivative, and $\langle \dots \rangle$ the ensemble average with respect to the noise, if present.
The infinite-time average velocity  is obtained by taking the limit
\begin{equation}
\label{LTAVdefinition}
\mathbf{\bar{V}}=\lim_{T_{\mathrm{e}}\to +\infty}\mathbf{\bar{v}}^{t_0,t_0+T_{\mathrm{e}}}\,.
\end{equation}
In most dissipative systems this limit exists and is independent of the initial time $t_0$ and of the initial conditions. 

In simulations, or real experiments, the variable
$\mathbf{r}(t)$ is known only in a finite time-interval.  Thus, the limit in Eq.~(\ref{LTAVdefinition}) can be calculated only approximately by choosing a sufficiently large value for the elapsed time $T_{\mathrm{e}}$. To be more precise, $T_{\mathrm{e}}$ must be chosen to be much larger than all other relevant time-scales in the problem. In particular, if the system is subjected to a periodic time-dependent forcing function with period $T$, then a necessary  condition is that $T_{\mathrm{e}} \gg T$.  

The main focus of the present work is to determine scaling properties with the finite-time $T_{\mathrm{e}}$. Our results will be applicable to any system, with a well-defined infinite-time average~(\ref{LTAVdefinition}), containing time-dependent forcing functions of the form
\begin{equation}
\label{forcingdef}
\mathbf{F}(t)= \boldsymbol{\Phi}\left(\omega_1 t, \omega_2 t\right)\,.
\end{equation}
In the above expression, $\boldsymbol{\Phi}\left(\theta_1,\theta_2\right)$ is a vector function~\cite{footnote1} periodic with respect to both arguments with period $2 \pi$ [i.e., $\boldsymbol{\Phi}\left(\theta_1+2\pi,\theta_2\right)=\boldsymbol{\Phi}\left(\theta_1,\theta_2+2\pi\right)=\boldsymbol{\Phi}\left(\theta_1,\theta_2\right)$ $\forall (\theta_1,\theta_2)\in \mathbb{R}^2$], and $\omega_1$ and $\omega_2$ are two positive parameters with the dimensions of frequency. Bifrequency forcing functions of this type have been extensively used in the literature  (e.g.,~\cite{Optical_Trap3,Bifrequency,Optical_Trap4}).  If  $\omega_1$ and $\omega_2$ are commensurate, i.e.,  $\omega_2=p \omega_1/q$ with $p$ and $q$  coprimes,  $\mathbf{F}(t)$ is periodic  of period $T=2\pi q/\omega_1$.

 To be more specific, we consider the following situation: one of the frequencies, say $\omega_1$, is kept fixed, while the other frequency, $\omega_2$, is scanned around a fixed value, $\omega_{2,0}$, which is commensurate with $\omega_1$. This corresponds to  quantum-optical setups of recent experiments \cite{Optical_Trap3}. We will indicate the ratio between 
$\omega_{2,0}$ and $\omega_1$  with $\omega_{2,0}/\omega_1=p_0/q_0$, with $p_0$ and $q_0$ coprimes. We aim to derive the general asymptotic  properties of the function $\mathbf{\bar{v}}^{t_0,t_0+T_{\mathrm{e}}} (\omega_2)$ in an interval around $\omega_{2,0}$. As a first step, we analyze to which extent the infinite-time average $\mathbf{\bar{V}}(\omega_2)$ can be approximated by the finite-time average  $\mathbf{\bar{v}}^{t_0,t_0+T_{\mathrm{e}}}(\omega_2)$. This analysis will reveal a general scaling property which will allow us to derive the asymptotic limit.

As discussed above, a necessary condition to approximate  the infinite-time limit  $\mathbf{\bar{V}}(\omega_{2,0})$  by the finite-time average $\mathbf{\bar{v}}^{t_0,t_0+T_{\mathrm{e}}}(\omega_{2,0})$ is that $T_{\mathrm{e}}\gg T_0=2\pi q_0/\omega_1$; i.e., the elapsed time should be significantly larger than the period of the driving. However, as we now show, no matter how large $T_{\mathrm{e}}$ is chosen, there always exists a frequency interval centered at $\omega_{2,0}$ within which it is not appropriate
 to approximate the average velocity $\mathbf{\bar{V}}(\omega_2)$ by the finite-time average $\mathbf{\bar{v}}^{t_0,t_0+T_{\mathrm{e}}}(\omega_2)$, except for the value $\omega_2=\omega_{2,0}$ itself.
Indeed, let us consider the frequency interval defined by the inequality $\left|\omega_2- \omega_{2,0}\right|<2\pi/(q_0 T_{\mathrm{e}})$,  and choose any value $\omega_{2,1}\neq\omega_{2,0}$ in this interval of the form $\omega_{2,1}=p_1\omega_1/q_1$, with $p_1$ and $q_1$ being coprime integers. Taking into account that there are infinitely many rational numbers within any real interval of nonzero length, the number of possible choices for $\omega_{2,1}$ is infinity. Since $|p_1 q_0-p_0 q_1|\geq 1$, it follows that $\omega_1/(q_0 q_1)\leq |\omega_{2,1}-\omega_{2,0}|<2\pi/(q_0 T_{\mathrm{e}})$. Thus, $T_1=2 \pi q_1/\omega_1>T_{\mathrm{e}}$ and, consequently, for the chosen value of $T_{\mathrm{e}}$,  it is not appropriate to approximate  
$\mathbf{\bar{V}}(\omega_{2,1})$ by $\mathbf{\bar{v}}^{t_0, t_0+T_{\mathrm{e}}}(\omega_{2,1})$. The same conclusion can be drawn if $\omega_{2,1}/\omega_1$ is irrational, since any irrational number can be approximated to any degree of accuracy by rational numbers.

The above result shows that, given an averaging time $T_{\mathrm{e}}$, there is a frequency interval of length $ 4\pi/(q_0 T_{\mathrm{e}})$ centered at $\omega_{2,0}$ in which the behavior of the infinite-time limit $\mathbf{\bar{V}}(\omega_2)$ cannot be inferred from that of the finite-time average $\mathbf{\bar{v}}^{t_0,t_0+T_{\mathrm{e}}}(\omega_2)$. It is precisely the scaling with $T_\mathbf{e}$ of the length of such an interval which allows us to derive the asymptotic behavior of the function $\mathbf{\bar{v}}^{t_0,t_0+T_{\mathrm{e}}}(\omega_2)$, the quantity that is directly measurable in a simulation or experiment, within the above-mentioned range of $\omega_2$ values.

\section{Derivation of the asymptotic properties}

\label{DOAP}

In this section, we analyze the asymptotic behavior of the average velocity $\mathbf{\bar{v}}^{t_0,t_0+T_{\mathrm{e}}}(\omega_2)$ for $\omega_2$ values such that $\left|\omega_2- \omega_{2,0}\right|$ is of the same order of magnitude as $ 2\pi/(q_0 T_{\mathrm{e}})$, and sufficiently large values of $T_{\mathrm{e}}$. To this purpose, we make a change of variables from the variable $\delta\omega_2=\omega_2- \omega_{2,0}$ to a dimensionless variable $\Delta\tilde{\omega}_2$, which is expressed in terms of $\delta \omega_2$ according to $\Delta\tilde{\omega}_2=q_0 \delta \omega_2 T_{\mathrm{e}}/(2 \pi)$. The average velocity as a function of this new variable will be given by $\boldsymbol{\bar{\nu}}_{\omega_{2,0}}^{t_0, t_0+T_{\mathrm{e}}}(\Delta\tilde{\omega}_2)=\mathbf{\bar{v}}^{t_0, t_0+T_{\mathrm{e}}}\left[\omega_{2,0}+2 \pi\Delta\tilde{\omega}_2/(q_0 T_{\mathrm{e}})\right]$, where the subscript $\omega_{2,0}$ indicates the midpoint of the region under study. In order to enforce the correct scaling derived in the previous section, we will consider the limit as $T_{\mathrm{{e}}}\rightarrow\infty$ and $\delta \omega_2\rightarrow 0$, so that $\Delta\tilde{\omega}_2$ is held constant.
Then, the leading-order of the aforesaid asymptotic behavior can be obtained from the limit 
\begin{equation}
\label{limit2}
\boldsymbol{\bar{\mathcal{V}}}_{\omega_{2,0}}(\Delta\tilde{\omega}_2)=\lim_{T_{\mathrm{e}}\to +\infty}\boldsymbol{\bar{\nu}}_{\omega_{2,0}}^{t_0, t_0+T_{\mathrm{e}}}(\Delta\tilde{\omega}_2)\,,
\end{equation}
while keeping $\Delta\tilde{\omega}_2$ fixed. As a fact useful in the following, we notice that the time-dependent forcing function in Eq.~(\ref{forcingdef}) can be rewritten in the form
\begin{equation}
\label{property1}
\mathbf{F}(t)=\boldsymbol{\Phi}\left[\omega_1 t, \omega_{2,0} t +\varphi(t)\right]\,,
\end{equation}
where
${ \varphi(t)= \delta \omega_2 t =2\pi \Delta\tilde{\omega}_2 t/(q_0 T_{\mathrm{e}})}$.

In order to find a more explicit expression for the asymptotic result $\boldsymbol{\bar{\mathcal{V}}}_{\omega_{2,0}}(\Delta\tilde{\omega}_2)$, it is convenient to introduce an auxiliary function of $T_{\mathrm{e}}$, $N(T_{\mathrm{e}})$, with the following three properties~\cite{footnote2}: (i) $N(T_{\mathrm{e}})$ is  dimensionless, integer, and positive for all $T_{\mathrm{e}}>0$, (ii) $\lim_{T_{\mathrm{e}}\to +\infty}N(T_{\mathrm{e}})=+\infty$, and (iii) $\lim_{T_{\mathrm{e}}\to +\infty}N(T_{\mathrm{e}})/T_{\mathrm{e}}=0$. By using this function, one can divide the interval $[t_0,t_{\mathrm{f}}]$ into $N(T_{\mathrm{e}})$ subintervals of equal length $\Delta \tau(T_{\mathrm{e}})=T_{\mathrm{e}}/N(T_{\mathrm{e}})$. Then, taking into account the definition in Eq.~(\ref{AVdefinition}), it is possible to rewrite Eq.~(\ref{limit2}) as
\begin{equation}
\label{limit3}
\boldsymbol{\bar{\mathcal{V}}}_{\omega_{2,0}}(\Delta\tilde{\omega}_2)=\lim_{T_{\mathrm{e}}\to +\infty}\frac{\Delta \tau(T_{\mathrm{e}})}{T_{\mathrm{e}}}\sum_{j=1}^{N(T_{\mathrm{e}})}
\boldsymbol{\bar{\nu}}_{\omega_{2,0}}^{\tau_{j-1},\tau_j}(\Delta\tilde{\omega}_2)\,,
\end{equation}
where $\tau_j=\tau_j(T_{\mathrm{e}})=t_0+j \Delta \tau(T_{\mathrm{e}})$. 

We now use two properties that follow from the definition of the function $N(T_{\mathrm{e}})$. First, from (iii) it follows that $\lim_{T_{\mathrm{e}}\to +\infty}\Delta \tau(T_{\mathrm{e}})=+\infty$, i.e., the length of the above-mentioned subintervals  tends to infinity as $T_{\mathrm{e}}\to +\infty$.  Second, from (ii) it follows that $\lim_{T_{\mathrm{e}}\to +\infty}\Delta\varphi (T_{\mathrm{e}})=0$, where $\Delta\varphi (T_{\mathrm{e}})=2\pi \Delta\tilde{\omega}_2 \Delta \tau(T_{\mathrm{e}})/(q_0 T_{\mathrm{e}})$ is the increment in $\varphi(t)$ corresponding to an increment $\Delta \tau(T_{\mathrm{e}})$ in $t$.  In the calculation of this last limit, the variable $\Delta\tilde{\omega}_2$ is held constant.
Thus, within each of these subintervals the function $\varphi(t)$ in Eq.~(\ref{property1}) can be approximated by a constant  (for instance, by its value at the left endpoint of the subinterval). This last approximation becomes exact in the limit $T_{\mathrm{e}}\to +\infty$. From these two properties and Eq.~(\ref{LTAVdefinition}), it follows that 
\begin{equation}
\label{limit4}
\boldsymbol{\bar{\mathcal{V}}}_{\omega_{2,0}}(\Delta\tilde{\omega}_2)=\frac{q_0}{2\pi \Delta\tilde{\omega}_2}\lim_{T_{\mathrm{e}}\to +\infty}\sum_{j=1}^{N(T_{\mathrm{e}})}\mathbf{\bar{V}}(\omega_{2,0},\varphi_{j-1})\Delta \varphi\,,
\end{equation}
where $\Delta \varphi=\Delta \varphi(T_{\mathrm{e}})$, $\varphi_j=\varphi_j(T_{\mathrm{e}})=\varphi[\tau_j(T_{\mathrm{e}})]=
2\pi \Delta\tilde{\omega}_2 t_0/(q_0 T_{\mathrm{e}})+j \Delta \varphi(T_{\mathrm{e}})$, and $\mathbf{\bar{V}}(\omega_{2,0},\varphi_{j-1})$ is the infinite-time average velocity corresponding to a forcing function of the form (\ref{property1}) but with a time-independent phase shift equal to $\varphi_{j-1}$. Finally, taking into account that the sum in Eq.~(\ref{limit4}) is a Riemann sum, as well as,  that $\lim_{T_{\mathrm{e}}\to+\infty}\varphi_0(T_{\mathrm{e}})=0$ and $\lim_{T_{\mathrm{e}}\to+\infty}\varphi_{N(T_{\mathrm{e}})}(T_{\mathrm{e}})=2\pi \Delta\tilde{\omega}_2/q_0$, one obtains
\begin{equation}
\label{res1}
\boldsymbol{\bar{\mathcal{V}}}_{\omega_{2,0}}(\Delta\tilde{\omega}_2)=\frac{q_0}{2\pi \Delta\tilde{\omega}_2}\int_{0}^{2\pi \Delta\tilde{\omega}_2/q_0}\mathrm{d}\varphi
\,\mathbf{\bar{V}}(\omega_{2,0},\varphi)\,.
\end{equation}
This expression is one of the central results of the present work. It shows that the asymptotic behavior $\boldsymbol{\bar{\mathcal{V}}}_{\omega_{2,0}}(\Delta\tilde{\omega}_2)$ is completely determined by $\mathbf{\bar{V}}(\omega_{2,0},\varphi)$, that is, by the infinite-time average velocity corresponding to a forcing function $\mathbf{F}_0(t, \varphi)=\boldsymbol{\Phi}\left(\omega_1 t, \omega_{2,0} t +\varphi\right)$, considered as a function of the time-independent phase shift $\varphi$.

Equation~(\ref{res1}) allows us to draw conclusions about the behavior of the asymptotic result $\boldsymbol{\bar{\mathcal{V}}}_{\omega_{2,0}}(\Delta\tilde{\omega}_2)$ from the specific features of the function 
$\mathbf{\bar{V}}(\omega_{2,0},\varphi)$. For instance, it is easy to show that, under quite general assumptions, $\mathbf{\bar{V}}(\omega_{2,0},\varphi)$ is a periodic function in $\varphi$ with period $2 \pi /q_0$; i.e., 
\begin{equation}
\label{periodicity0}
\mathbf{\bar{V}}(\omega_{2,0},\varphi)=\mathbf{\bar{V}}(\omega_{2,0},\varphi+2 \pi/q_0)
\end{equation}
for all $\varphi$. Indeed, since $p_0$ and $q_0$ are coprime,  from  the B\'ezout identity~\cite{Bezout} it follows that there exist integers $l$ and $n$ such that $l p_0+n q_0=1$. From this equality, one can readily check that $ \mathbf{F}_0(t+2 \pi l/\omega_1,\varphi)=\mathbf{F}_0(t,\varphi+2\pi/q_0)$. Therefore, by assuming that all the time-dependent noise sources (if any) can be described by stationary stochastic processes, one has that $\mathbf{\bar{v}}^{t_0^{\prime},t_0^{\prime}+T_{\mathrm{e}}}(\omega_{2,0},\varphi)=\mathbf{\bar{v}}^{t_0,t_0+T_{\mathrm{e}}}(\omega_{2,0},\varphi+2\pi/q_0)$, with $t_0^{\prime}=t_0+2 \pi l/\omega_1$.   Equation~(\ref{periodicity0}) then follows
from the independence of the limit in Eq.~(\ref{LTAVdefinition}) on the initial conditions. Notice that, according to Eq.~(\ref{res1}), the average of $\mathbf{\bar{V}}(\omega_{2,0},\varphi)$ over a period $2\pi/q_0$ is simply given by $\boldsymbol{\bar{\mathcal{V}}}_{\omega_{2,0}}(1)$. Taking into account Eqs.~(\ref{res1}) and (\ref{periodicity0}), it is easy to verify that
\begin{equation}
\label{zeros1}
\boldsymbol{\bar{\mathcal{V}}}_{\omega_{2,0}}(k)=\boldsymbol{\bar{\mathcal{V}}}_{\omega_{2,0}}(1)
\end{equation}
for any nonzero integer $k$. Similarly, one can check that
\begin{equation}
\label{antisymmetry1}
\boldsymbol{\bar{\mathcal{V}}}_{\omega_{2,0}}\left(-k/2\right)-\boldsymbol{\bar{\mathcal{V}}}_{\omega_{2,0}}(1)=\boldsymbol{\bar{\mathcal{V}}}_{\omega_{2,0}}(1)-\boldsymbol{\bar{\mathcal{V}}}_{\omega_{2,0}}
\left(k/2\right)
\end{equation}
for any odd integer $k$. Equations~(\ref{zeros1}) and (\ref{antisymmetry1}) reveal two remarkable properties of the asymptotic result $\boldsymbol{\bar{\mathcal{V}}}_{\omega_{2,0}}(\Delta\tilde{\omega}_2)$ which follow directly from the periodicity of the infinite-time average velocity expressed in Eq.~(\ref{periodicity0}).

In order to gain some insight into the qualitative features of the function $\boldsymbol{\bar{\mathcal{V}}}_{\omega_{2,0}}(\Delta\tilde{\omega}_2)$, let us consider first its behavior in the vicinity of the point $\Delta\tilde{\omega}_2=0$. From Eqs.~(\ref{limit2}) and (\ref{LTAVdefinition}), it is clear that 
$\boldsymbol{\bar{\mathcal{V}}}_{\omega_{2,0}}(0)=\mathbf{\bar{V}}(\omega_{2,0},0)$. Furthermore, by applying L'H\^opital's rule, one can easily show from Eq.~(\ref{res1}) that $\lim_{\Delta\tilde{\omega}_2\to 0} \boldsymbol{\bar{\mathcal{V}}}_{\omega_{2,0}}(\Delta\tilde{\omega}_2)=\mathbf{\bar{V}}(\omega_{2,0},0)$. Thus,  $\boldsymbol{\bar{\mathcal{V}}}_{\omega_{2,0}}(\Delta\tilde{\omega}_2)$ is continuous at $\Delta\tilde{\omega}_2=0$ and approaches the value $\mathbf{\bar{V}}(\omega_{2,0},0)$ as $\Delta\tilde{\omega}_2$ tends to zero. To understand what happens when $|\Delta\tilde{\omega}_2|$ increases,
it is convenient to introduce the function
$\boldsymbol{\bar{\mathcal{D}}}_{\omega_{2,0}}(\Delta\tilde{\omega}_2)=2 \pi \Delta\tilde{\omega}_2\left[ \boldsymbol{\bar{\mathcal{V}}}_{\omega_{2,0}}(\Delta\tilde{\omega}_2)-\boldsymbol{\bar{\mathcal{V}}}_{\omega_{2,0}}(1)\right]/q_0$,  so that 
\begin{equation}
\label{res1new}
\boldsymbol{\bar{\mathcal{V}}}_{\omega_{2,0}}(\Delta\tilde{\omega}_2)=\boldsymbol{\bar{\mathcal{V}}}_{\omega_{2,0}}(1)+\frac{q_0 \,\boldsymbol{\bar{\mathcal{D}}}_{\omega_{2,0}}(\Delta\tilde{\omega}_2)}{2\pi \Delta\tilde{\omega}_2}\,.
\end{equation}
 From the definition of  $\boldsymbol{\bar{\mathcal{D}}}_{\omega_{2,0}}(\Delta\tilde{\omega}_2)$ and Eq.~(\ref{res1}), it is straightforward to see that $\boldsymbol{\bar{\mathcal{D}}}_{\omega_{2,0}}(\Delta\tilde{\omega}_2)=\int_{0}^{2\pi \Delta\tilde{\omega}_2/q_0}\mathrm{d}\varphi
\,\left[ \mathbf{\bar{V}}(\omega_{2,0},\varphi)-\boldsymbol{\bar{\mathcal{V}}}_{\omega_{2,0}}(1)\right]$. Therefore, given that $\mathbf{\bar{V}}(\omega_{2,0},\varphi)$ is a periodic function in $\varphi$ with period $2 \pi /q_0$ and average value $\boldsymbol{\bar{\mathcal{V}}}_{\omega_{2,0}}(1)$, it follows  that $\boldsymbol{\bar{\mathcal{D}}}_{\omega_{2,0}}(\Delta\tilde{\omega}_2)$ is a periodic function of $\Delta\tilde{\omega}_2$ with period $1$. Consequently, according to Eq.~(\ref{res1new}), the asymptotic result $\boldsymbol{\bar{\mathcal{V}}}_{\omega_{2,0}}(\Delta\tilde{\omega}_2)$ approaches the value $\boldsymbol{\bar{\mathcal{V}}}_{\omega_{2,0}}(1)$ in an oscillatory manner as $|\Delta\tilde{\omega}_2|$ increases, i.e., as $\omega_2$ moves away from the value $\omega_{2,0}$. The result given in Eq.~(\ref{zeros1}) is a further indication of this oscillatory behavior.

\section{ Model-specific examples}

\label{MSE}

To illustrate our results, we will use a one-dimensional model consisting of a Brownian particle, with mass $m$ and position $x$, moving in a periodic potential $U(x)$, with period $\lambda$,  under the influence of a biharmonic force of the form $F(t)=A\left[\cos(\omega_1 t)+\cos(\omega_2 t)\right]$. The dynamics of this system is described by the stochastic differential equation 
\begin{equation}
\label{firstdynamics}
m \,\ddot{x}(t)=-\,\alpha \,\dot{x}(t)-U^{\prime}\left[x(t)\right]+F(t)+\sqrt{2 \Gamma}\,\xi(t)\,,
\end{equation}
where $U^{\prime}(x)$ is the derivative with respect to $x$ of $U(x)$, $\alpha$ is the friction coefficient, $\xi(t)$ is a Gaussian white noise of zero mean and autocorrelation $\langle \xi(t)\xi(s)\rangle =\delta(t-s)$, and $\Gamma$ is the noise strength. The periodic potential $U(x)$ can always be expressed as $U(x)=U_0 \, \Psi(2\pi x/\lambda)/2$, where $U_0$ is a constant with the dimensions of energy and $\Psi(\theta)$ is a dimensionless periodic function with period $2 \pi$. By expressing Eq.~(\ref{firstdynamics}) in terms of the dimensionless quantities $\hat{t}=\omega_1 t$ and $\hat{x}=\pi x/\lambda$, the  model is completely determined by five dimensionless parameters, namely, $\hat{m}=m \,\omega_1/\alpha$, $\hat{U}_0=\pi^2 U_0/(\lambda^2\alpha\,\omega_1)$, $\hat{A}=\pi A/(\lambda\, \alpha\,\omega_1)$, $\hat{\Gamma}=\pi^2 \Gamma/(\lambda^2  \alpha^2 \omega_1)$, and $\hat{\omega}_2=\omega_2/\omega_1$. 

\begin{figure}
\begin{center}
\includegraphics[scale=0.8]{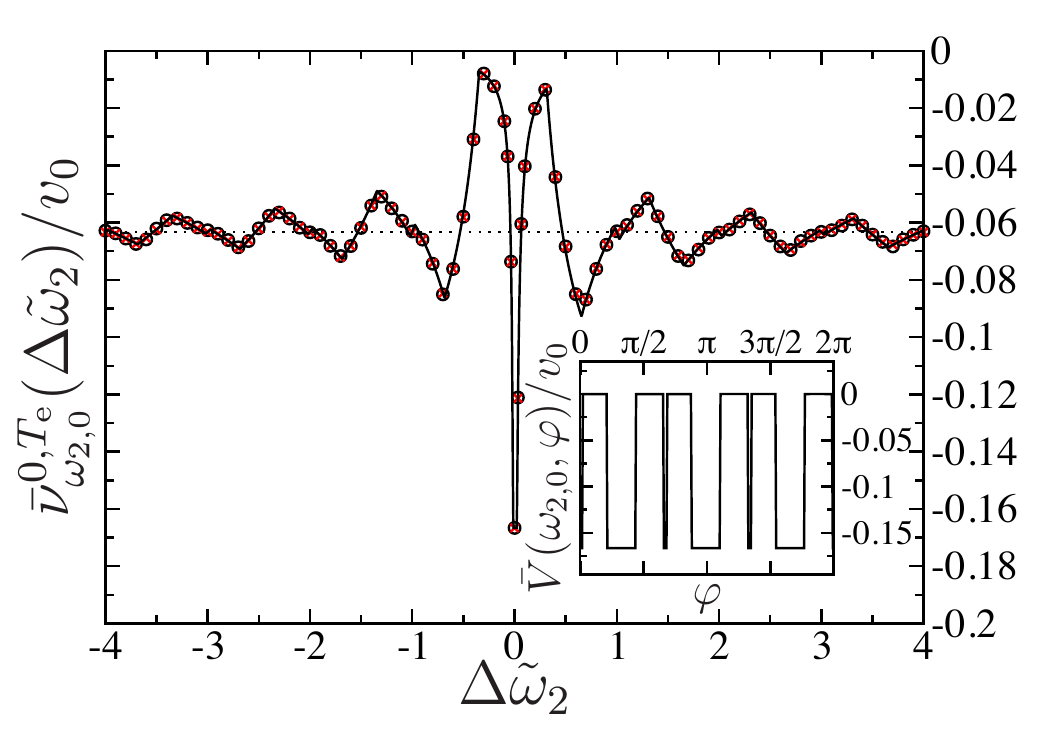}
\end{center}
\caption{(Color online) Dependence of the dimensionless average velocity $\bar{\nu}_{\omega_{2,0}}^{0, T_{\mathrm{e}}}(\Delta\tilde{\omega}_2)/v_0$ on the dimensionless variable $\Delta \tilde{\omega}_2$ for the model given by Eq.~(\ref{firstdynamics}), with $U(x)=U_0 \left[\cos(2 \pi x/\lambda)+1/4 \cos(4 \pi x/\lambda+\pi/3)\right]/2$. The value $\Delta \tilde{\omega}_2=0$ corresponds to $\hat{\omega}_2=\hat{\omega}_{2,0}=p_0/q_0$, with $p_0=2$ and $q_0=3$. The other parameters are $\hat{m}=\hat{\Gamma}=0$, $\hat{U}_0=\hat{A}=1$, $T_{\mathrm{e}}=10^4\omega_1^{-1}$ (circles) and $T_{\mathrm{e}}=10^5\omega_1^{-1}$ (red crosses). The solid line indicates the asymptotic result $\bar{\mathcal{V}}_{\omega_{2,0}}(\Delta\tilde{\omega}_2)/v_0$ obtained by numerically evaluating the integral in Eq.~(\ref{res1}). The dotted line shows the value $\bar{\mathcal{V}}_{\omega_{2,0}}(1)/v_0\approx -0.0633$. The inset depicts the dimensionless infinite-time average velocity $\bar{V}(\omega_{2,0},\varphi)/v_0$ as a function of the phase shift $\varphi$. The infinite-time limit has been approximately calculated by evaluating the average velocity $\bar{v}^{0,T_{\mathrm{e}}}(\omega_{2,0},\varphi)$ for $T_{\mathrm{e}}=10^5 \omega_1^{-1}$.}
\label{fig1}
\end{figure}

In Fig.~\ref{fig1} we depict the dependence of the dimensionless average velocity  $\bar{\nu}_{\omega_{2,0}}^{0,T_{\mathrm{e}}}(\Delta\tilde{\omega}_2)/v_0$ on the variable $\Delta \tilde{\omega}_2$, where we have introduced the typical velocity $v_0=\lambda\, \omega_1/\pi$. Since the model is one-dimensional, the
boldface vector notation has been dropped. For the sake of simplicity, in this first figure we have considered the overdamped deterministic case  ($\hat{m}=\hat{\Gamma}=0$). Furthermore, in order to avoid the presence of symmetries which are discussed later, we have chosen a highly asymmetric periodic potential (see figure caption for details). The results shown in Fig.~\ref{fig1} clearly demonstrate that, for the elapsed time values considered (i.e., for $T_{\mathrm{e}}=10^4 \omega_1^{-1}$ and $T_{\mathrm{e}}=10^5 \omega_1^{-1}$), the average velocity $\bar{\nu}_{\omega_{2,0}}^{0, T_{\mathrm{e}}}(\Delta\tilde{\omega}_2)$ has essentially reached its asymptotic limit $\bar{\mathcal{V}}_{\omega_{2,0}}(\Delta\tilde{\omega}_2)$. Consequently, $\bar{\nu}_{\omega_{2,0}}^{0, T_{\mathrm{e}}}(\Delta\tilde{\omega}_2)$ displays the typical oscillatory behavior predicted by Eq.~(\ref{res1new}), and satisfies the conditions imposed by Eqs.~(\ref{zeros1}) and (\ref{antisymmetry1}). It is to be noted that, in this figure, $\bar{\mathcal{V}}_{\omega_{2,0}}(0)=\bar{V}(\omega_{2,0},0)=\lambda/T_0$, where $T_0=2 \pi q_0/\omega_1= 6\pi/\omega_1$ is the period of the biharmonic force (see Ref.~\cite{Ajdari} for an explanation of this fact).

We return now to the general case to investigate how the specific symmetries of the evolution equation affect the shape of $\boldsymbol{\bar{\mathcal{V}}}_{\omega_{2,0}}(\Delta\tilde{\omega}_2)$. As a first example, let us assume that  the differential equation describing the evolution of $\mathbf{r}(t)$ in the presence of the forcing function $\mathbf{F}_0(t, \varphi)$ is invariant under the transformation
${S}_1: \{\mathbf{r},\,\dot{\mathbf{r}},\,\ddot{\mathbf{r}},\,t,\,\varphi\}\rightarrow\left\{\mathbf{r}_1-\mathbf{r},\,-\dot{\mathbf{r}},\,-\ddot{\mathbf{r}},\,t+t_1,\,\varphi-\pi/q_0\right\}$,
where $\mathbf{r}_1$ and $t_1$ are constant parameters. Then, if $\mathbf{r}(t)$ is a solution of this differential equation corresponding to a given value $\varphi$ of the phase shift,  $\mathbf{R}(t)=\mathbf{r}_1-\mathbf{r}(t+t_1)$ is also a solution corresponding to the displaced value $\varphi+\pi/q_0$. Since $\langle \dot{\mathbf{R}}(t)\rangle=-\langle \dot{\mathbf{r}}(t+t_1)\rangle$ for all $t$, it follows from Eq.~(\ref{LTAVdefinition}) that
$\mathbf{\bar{V}}(\omega_{2,0},\varphi)=-\mathbf{\bar{V}}(\omega_{2,0},\varphi+\pi/q_0)$ for all $\varphi$. Thus,  from Eq.~(\ref{res1}), it is clear that $\boldsymbol{\bar{\mathcal{V}}}_{\omega_{2,0}}(1)=0$. Consequently, in this case, Eqs.~(\ref{zeros1}) and (\ref{antisymmetry1}) can be rewritten, respectively, as $\boldsymbol{\bar{\mathcal{V}}}_{\omega_{2,0}}(k)=0$ for any nonzero integer $k$, and
$\boldsymbol{\bar{\mathcal{V}}}_{\omega_{2,0}}\left(-k/2\right)=-\,\boldsymbol{\bar{\mathcal{V}}}_{\omega_{2,0}}\left(k/2\right)$
for any odd integer $k$. 
 According to the first of these last two equations it follows that, assuming that there exists a directed motion for $\omega_2=\omega_{2,0}$, this current can be suppressed by slightly shifting the frequency $\omega_2$ from the value $\omega_{2,0}$ to the nearby values $\omega_{2,0}\pm 2\pi /(q_0 T_{\mathrm{e}})$. This offers the possibility to discriminate two neighboring values of the frequency $\omega_2$ differing in a quantity $2\pi /(q_0 T_{\mathrm{e}})$, which is $q_0$ times smaller than the usually assumed Fourier width $2\pi/T_{\mathrm{e}}$ \cite{lille}.

As a second example, let us assume now that the aforesaid differential equation is invariant under the transformation 
$
S_2: \{\mathbf{r},\,\dot{\mathbf{r}},\,\ddot{\mathbf{r}},\,t,\,\varphi\}\rightarrow\left\{\mathbf{r}_2-\mathbf{r},\,\dot{\mathbf{r}},\,-\ddot{\mathbf{r}},\,t_2-t,\,-\varphi\right\}
$,
with $\mathbf{r}_2$ and $t_2$ being constants. Then, if $\mathbf{r}(t)$ is a solution corresponding to a given value $\varphi$ of the phase shift, $\mathbf{R}(t)=\mathbf{r}_2-\mathbf{r}(t_2-t)$ is also a solution corresponding to the opposite value $-\varphi$. Thus, following a similar reasoning as above, it is easy to show that $\mathbf{\bar{V}}(\omega_{2,0},-\varphi)=\mathbf{\bar{V}}(\omega_{2,0},\varphi)$ for all $\varphi$. Consequently, from Eq.~(\ref{res1}) it readily follows that $\boldsymbol{\bar{\mathcal{V}}}_{\omega_{2,0}}(\Delta\tilde{\omega}_2)$ is an even function of $\Delta\tilde{\omega}_2$; i.e.,  $\boldsymbol{\bar{\mathcal{V}}}_{\omega_{2,0}}(-\Delta\tilde{\omega}_2)=\boldsymbol{\bar{\mathcal{V}}}_{\omega_{2,0}}(\Delta\tilde{\omega}_2)$ for all $\Delta\tilde{\omega}_2$. In this case, Eq.~(\ref{antisymmetry1}) can be rewritten in the form $\boldsymbol{\bar{\mathcal{V}}}_{\omega_{2,0}}\left(k/2\right)=\boldsymbol{\bar{\mathcal{V}}}_{\omega_{2,0}}(1)$ for any odd integer $k$ or, if the differential equation is also invariant under the transformation $S_1$, in the form $\boldsymbol{\bar{\mathcal{V}}}_{\omega_{2,0}}\left(k/2\right)=0$. 

\begin{figure}
\begin{center}
\includegraphics[scale=0.80]{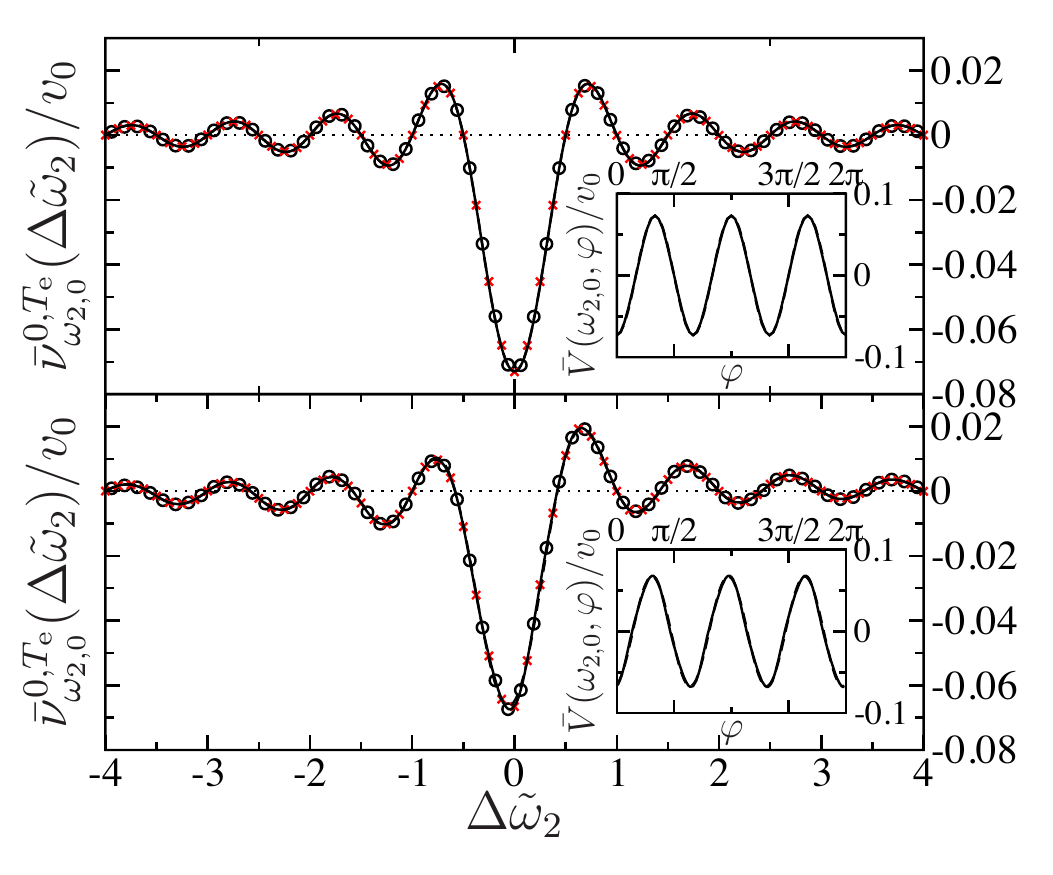}
\end{center}
\caption{(Color online) The same as in Fig.~\ref{fig1} but for a symmetric potential $U(x)=U_0 \cos(2\pi x/\lambda)/2$, a dimensionless noise strength $\hat{\Gamma}=0.05$, and two values of the dimensionless mass, namely, $\hat{m}=0$ (top panel) and $\hat{m}=1$ (bottom panel).
The rest of the parameters are the same as in Fig.~\ref{fig1}. The dashed lines indicate the results obtained from Eqs.~(\ref{cosform}) (insets) and (\ref{approx}) (main figures). The model-dependent
parameters are $\chi=\pi/3$ and $\bar{V}(\omega_{2,0},\chi)/v_0=0.073$ (top panel), and  $\chi=0.963$ and $\bar{V}(\omega_{2,0},\chi)/v_0=0.067$ (bottom panel). These dashed lines are indistinguishable from the corresponding solid lines.
The dotted lines show the value $\bar{\mathcal{V}}_{\omega_{2,0}}(1)/v_0=0$.}
\label{fig2}
\end{figure}

In Fig.~\ref{fig2} we depict again the dependence of the dimensionless average velocity  $\bar{\nu}_{\omega_{2,0}}^{0,T_{\mathrm{e}}}(\Delta\tilde{\omega}_2)/v_0$ on the variable $\Delta \tilde{\omega}_2$ for the same model as in Fig.~\ref{fig1}, given by Eq.~(\ref{firstdynamics}), but with a symmetric potential and a nonzero noise strength (see figure caption for details). We consider both the overdamped case (top panel) and the underdamped case (bottom panel). In both cases, it is easy to see that the corresponding evolution equations are invariant under the one-dimensional version of the transformation~$S_1$ with, e.g., $t_1=-\pi/\omega_1$ and $x_1=0$ \cite{footnote3}.  Consequently, in both panels $\bar{\mathcal{V}}_{\omega_{2,0}}(1)=0$ (see the dotted lines). However, only in the case of the overdamped dynamics, there exist values for $t_2$ and $x_2$ such that the evolution equation is invariant under the one-dimensional version of the transformation~$S_2$,  e.g., $t_2=0$ and $x_2=\lambda/2$. Thus, only the top panel shows the features associated with this transformation.

At this point, it is important to remember that, in many one-dimensional systems driven by biharmonic forces,  $\bar{V}(\omega_{2,0},\varphi)$ can be expressed approximately as 
\begin{equation}
\label{cosform}
\bar{V}(\omega_{2,0},\varphi)=\bar{V}(\omega_{2,0},\chi) \cos\left[q_0(\varphi-\chi) \right]\,,
\end{equation}
where $\chi$ is the value of $\varphi$ in the interval $[0,2\pi/q_0)$ for which $\bar{V}(\omega_{2,0},\varphi)$ attains its maximum value $\bar{V}(\omega_{2,0},\chi)$. As shown in Ref.~\cite{Niurka1}, the functional form in Eq.~(\ref{cosform}) is a consequence solely of the system symmetries, being independent of the interaction details. By introducing Eq.~(\ref{cosform}) in Eq.~(\ref{res1}), one obtains 
\begin{equation}
\label{approx}
\bar{\mathcal{V}}_{\omega_{2,0}}(\Delta\tilde{\omega}_2)=\bar{V}(\omega_{2,0},\chi)\,\frac{\sin\left(2 \pi\Delta\tilde{\omega}_2 -q_0\chi \right)+\sin(q_0\chi)}{2 \pi \Delta\tilde{\omega}_2}\,.
\end{equation}
This expression is, in fact, a special case of Eq.~(\ref{res1new}) where $\bar{\mathcal{V}}_{\omega_{2,0}}(1)=0$ and $\bar{\mathcal{D}}_{\omega_{2,0}}(\Delta\tilde{\omega}_2)=\bar{V}(\omega_{2,0},\chi)\left[\sin\left(2 \pi\Delta\tilde{\omega}_2 -q_0\chi \right)+\sin(q_0\chi) \right]/q_0 $. 

The significance of Eq.~(\ref{approx}) lies in the fact that it is an explicit expression for the asymptotic result $\bar{\mathcal{V}}_{\omega_{2,0}}(\Delta\tilde{\omega}_2)$ with only two  model-dependent  parameters, namely, $\chi$ and $\bar{V}(\omega_{2,0},\chi)$. Particular cases of the sub-Fourier scaling  with $2\pi/(q_0 T_{\mathrm{e}})$ evidenced by Eq.~(\ref{approx}) have been proposed in the literature based on heuristic arguments~\cite{Optical_Trap3,dcubero1}, as well as observed in cold-atom experiments \cite{Optical_Trap3}.
The results obtained from Eqs.~(\ref{cosform}) and (\ref{approx}), with an appropriate choice of the model parameters, are indicated in Fig.~\ref{fig2} with dashed lines. The agreement is so good that it is nearly impossible to distinguish between the solid and the dashed lines. By contrast, a glance at the inset of Fig.~\ref{fig1} reveals that it is not reasonable to approximate that curve by a function of the form given by Eq.~(\ref{cosform}).

\section{Conclusions}

\label{Conclusions}

In conclusion, we determined a universal asymptotic behavior of the average of a variable over a finite time interval. Our main results, expressed by Eqs.~(\ref{res1})--(\ref{res1new}), are applicable to any system driven by a bifrequency forcing function and with a well-defined infinite-time average, independent of the initial conditions. With the help of Eq.~(\ref{res1}) and some symmetry considerations, we have obtained some qualitative and quantitative information about the aforesaid asymptotic behavior. In particular, we have derived an asymptotic expression for 
the width of the resonance observed by keeping one frequency fixed, and varying the other one. We have showed 
that this width is smaller than the one predicted by the Fourier inequality \cite{lille} by a 
factor determined by the two driving frequencies, and independent of the model system parameters.
The nonperturbative approach presented here to determine the asymptotic scaling finds direct application in the identification of non-linear systems displaying sub-Fourier resonances
\cite{lille,Optical_Trap3}.

\acknowledgements{We acknowledge financial support from the Leverhulme
Trust and the Ministerio de Ciencia e 
Innovaci\'on of Spain FIS2008-02873 (J. C.-P. and D. C.).}


\begin{thebibliography}{99}
\bibitem{Reviews}  L. Gammaitoni, P. H\"anggi, P. Jung, and F. Marchesoni, Rev. Mod. Phys. {\bf 70}, 223 (1998); R. D. Astumian and P. H\"anggi, Phys. Today {\bf 55}(11), 33 (2002); P. Reimann, Phys. Rep. {\bf 361} 57 (2002); P. H\"anggi, F. Marchesoni, and F. Nori, Ann. Phys. (Leipzig) {\bf 14}, 51 (2005); P. H\"anggi and F. Marchesoni, Rev. Mod. Phys. {\bf 81}, 387
(2009).
\bibitem{lille} P. Szriftgiser, J. Ringot, D. Delande, and J. C. Garreau, Phys. Rev. Lett. {\bf 89}, 224101 (2002).
\bibitem{Optical_Trap3}  R. Gommers, M. Brown, and F. Renzoni, Phys. Rev. A {\bf 75}, 053406 (2007).
\bibitem{footnote1} The number of components of the vector  $\boldsymbol{\Phi}\left(\theta_1,\theta_2\right)$ may not coincide with the spatial dimension of the model. For instance, in the one-dimensional system considered in Ref.~\cite{Optical_Trap4} it has two components.
\bibitem{Bifrequency} S. Flach, O. Yevtushenko, and Y. Zolotaryuk, Phys. Rev. Lett. {\bf 84}, 2358 (2000); O. Yevtushenko, S. Flach, Y. Zolotaryuk, and A. A. Ovchinnikov, Europhys. Lett. {\bf 54}, 141 (2001); M. Schiavoni, L. Sanchez-Palencia, F. Renzoni, and G. Grynberg, Phys. Rev. Lett. {\bf 90}, 094101 (2003); P. H. Jones, M. Goonasekera, and F. Renzoni, Phys. Rev. Lett. {\bf 93}, 073904 (2004); R. Gommers et al., Phys. Rev. Lett. {\bf 94}, 143001 (2005); R. Gommers, S. Bergamini, and
F. Renzoni, Phys. Rev. Lett. {\bf 95}, 073003 (2005); R. Gommers, S. Denisov, and F. Renzoni,
Phys. Rev. Lett. {\bf 96}, 240604 (2006);
 A. Wickenbrock et al., Phys. Rev. Lett. {\bf 108}, 020603 (2012).
\bibitem{Optical_Trap4} R. Gommers, V. Lebedev, M. Brown, and F. Renzoni, Phys. Rev. Lett. {\bf 100}, 040603 (2008).
\bibitem{footnote2} An example of such a function would be $N(T_{\mathrm{e}})=\lfloor \sqrt{T_{\mathrm{e}}/T_0}\rfloor$, with  $\lfloor z\rfloor$ being the floor function of $z$, i.e., the greatest integer less than or equal to $z$.
\bibitem{Bezout} G. A. Jones and J. M. Jones, {\it Elementary Number Theory} (Springer Verlag, London, 1998).
\bibitem{Ajdari} A. Ajdari, D. Mukamel, L. Peliti, and J. Prost, J. Phys.
I France {\bf 4}, 1551 (1994).
\bibitem{footnote3} Notice that the stochastic processes $\xi(t)$, $-\xi(t+t_1)$, and $\xi(t_2-t)$ are completely equivalent.
\bibitem{Niurka1}   N. R. Quintero, J. A. Cuesta, and R. Alvarez-Nodarse, Phys. Rev. E {\bf 81}, 030102(R) (2010).
\bibitem{dcubero1}  D. Cubero and F. Renzoni, Phys. Rev. E {\bf 86}, 056201 (2012).
\end{thebibliography}
\end{document}